\documentclass[11pt]{article}
\pdfoutput=1

\usepackage[T1]{fontenc}              % T1 font encoding as default
\usepackage[latin9]{inputenc}         % ISO-8859-15 (Latin 9) text

\usepackage[a4paper]{geometry}        % DIN A4 paper

\usepackage{url}

\usepackage{booktabs}
\usepackage{subfig}

\usepackage{graphicx} % WE USE PDFLATEX
\usepackage{amssymb,amsfonts,amsmath}

\usepackage{natbib}
\usepackage{mathpazo}
\usepackage{microtype}

\usepackage{color}
\newcommand\Fdr{\text{Fdr}}

\newcommand\fdr{\text{fdr}}

\newcommand\fdrtool{\texttt{fdrtool}\ }

\newcommand{\figcite}[1]{Fig.~\textbf{\ref{#1}}}
\newcommand{\eqcite}[1]{Eq.~\textbf{\ref{#1}}}
\newcommand{\tabcite}[1]{Tab.~\textbf{\ref{#1}}}

\begin{document}

\date{27 April 2011\\Accepted for publication in \\ Journal de la Soci\'et\'e Fran{\c c}aise de Statistique }

\title{
Learning False Discovery Rates By Fitting Sigmoidal Threshold Functions
}

\author{Bernd Klaus 
      \thanks{Institute for Medical Informatics,
      Statistics and Epidemiology,
      University of Leipzig,
      H\"artelstr. 16--18,
      D-04107 Leipzig, Germany} 
{} and
 Korbinian Strimmer \footnotemark[1]
}

\maketitle
\begin{abstract}

   False discovery rates  (FDR) are typically estimated from a mixture of a null
and an alternative distribution. Here, we study a complementary approach proposed by
Rice and Spiegelhalter (2008) that uses as primary quantities the null model and a parametric
family for the local false discovery rate.  Specifically, we consider the half-normal
decay  and the beta-uniform mixture models as FDR threshold functions.
Using simulations and analysis of real data we compare the           
performance of the Rice-Spiegelhalter approach with that of competing FDR estimation
procedures. If the alternative model is misspecified and an empirical null  distribution
is employed the accuracy of FDR estimation degrades substantially. Hence, 
while being a very elegant formalism, the FDR threshold approach requires special
care in actual application.

\end{abstract}

\newpage

\section{Introduction}

Statistical techniques for multiple testing have become
indispensable in the analysis of modern high-dimensional data \cite{Ben2010}.
One of the most prominent approaches uses false discovery rates (FDR)
as a measure of error and for determining test thresholds.
A precursor of the FDR approach was presented in 
\cite{SS82} but only with the seminal work of Benjamini and Hochberg \citep{BH95}
FDR was firmly established in the statistical community.

FDR estimation and control is best understood from a combined Bayesian-frequentist
perspective \citep{Efr08a}.  In this view, the data
are modeled by a two-component mixture and local FDR is defined as the Bayesian
posterior probability of the null model given the observed value of a test statistic.
In an interesting comment Rice and Spiegelhalter \cite{RS2008} reverse the traditional view
prevalent in FDR estimation.  Rather than assuming a null and alternative model
to derive FDR curves they proceed by specification of a null model plus a parametric family
for the local FDR threshold function.  The advantage of this procedure is that the alternative model 
needs not to be specified explicitly and that at the same time monotonicity of FDR is automatically enforced.

Here, we investigate the Rice-Spiegelhalter approach by studying two different choices of
threshold functions and considering two settings for the separation of null
and alternative mixture components. Using simulation and analysis of four data sets
we also provide a comparison with competing FDR estimation algorithms.

The remainder of the paper is as follows. First, we revisit the background for
FDR estimation using mixture models and local FDR threshold curves.  Next, we describe two simple models
for threshold curves, the beta-uniform mixture (BUM) and the half-normal decay (HND) model.
Subsequently, we study the performance of the approach both for a prespecified null model and for empirical null model estimation.

\section{False discovery rate estimation via threshold curves}

Estimation of false discovery rates (FDR) typically starts by fitting
a two-component mixture model  to the observed test statistics \citep{Efr08a}.
This mixture consists of a null
model $f_0$ and an alternative component $f_A$ from which the 
``interesting'' cases are assumed to be drawn. In the following we 
use a general test statistic $y \geq 0 $,
with large values of $y$ indicating an ``interesting'' and small values
close to zero an ``uninteresting'' case.  Examples for $y$ include
absolute $z$-scores $|z|$, absolute correlations $|r|$ or $1-p$, i.e the complement
of $p$-values. We can write the
mixture model in terms of densities as
$$
f(y)  =  \eta_0 f_0(y) + (1-\eta_0) f_A(y)
$$
and using distributions as
$$
F(y)  =  \eta_0 F_0(y) + (1-\eta_0) F_A(y) \, . 
$$
The parameter $\eta_0$ is the true proportion of the null features.
The statistic $y$ corresponds, e.g., to 1 - $p$-value or the absolute
value of a $z$-score or of a correlation coefficient \citep{Str08c}.
From a given mixture model the local FDR
(=fdr) is  obtained by 
\begin{equation}
\begin{split}
\label{eq:localfdr}
\fdr(y) & = \text{Prob}(\text{``not interesting''} | Y = y) \\   
&=  \eta_0 \frac{ f_0(y)}{f(y)}   \\
\end{split}  
\end{equation}
and the tail-area-based FDR (=Fdr), also known as $q$-value,
is defined by
\begin{equation}
\label{eq:tailareafdr}
\begin{split}
\text{Fdr}(y) & = \text{Prob}(\text{``not interesting''} | Y \geq y) \\   
&=  \eta_0 \frac{ 1-F_0(y)}{1-F(y)}  \,. \\
\end{split}  
\end{equation}
Most (if not all) proposed procedures for determining Fdr and fdr 
values can be characterized according to the strategies employed for estimation
of the underlying densities and distributions - 
for an overview see \citep{Str08c,Efr08a}.  For reasons of identifiability
of the mixture model 
the alternative component is assumed to vanish near the orgin and hence it follows that $\fdr(0)=1$.
Similarly, by construction we have $\text{Fdr}(0)=\eta_0$ as $F \rightarrow F_0$ for
small $y$. 

An alternative approach to FDR estimation is presented by \citep{RS2008}
who suggest to view the null model $f_0$ plus the fdr curve defined
by $\fdr(y)$ as the primary objects, rather than the two densities $f_0(y)$ and $f_A(y)$.
From \eqcite{eq:localfdr} we obtained the marginal distribution
\begin{equation}
\label{eq:marginaldist}
f(y) = \frac{\eta_0 f_0(y)}{ \fdr(y)} 
\end{equation}
which is here represented as a function of the null model and the local FDR.
Similarly, the alternative component is given by
$$
f_A(y) = \frac{\eta_0}{1-\eta_0} \frac{1 - \fdr(y)}{\fdr(y)}f_0(y) \,.
$$
Furthermore, as $f(y)$ is a density with $\int_0^\infty f(y) dy = 1$
we get the relationship
\begin{equation}
\label{eq:eta0integral}
\eta_0 = \biggl( \int_0^\infty \frac{f_0(y)}{\fdr(y)}dy \biggr)^{-1} \,.
\end{equation}
As a consequence, specifying $f_0(y)$ together with $\text{fdr}(y)$ 
is equivalent to the standard two-component formulation, but with 
$\eta_0$ and $f_A(y)$ viewed as derived rather than primary quantities.  
\eqcite{eq:marginaldist}
also plays an important role in the  \fdrtool
algorithm for FDR estimation (cf. \citep{Str08c}, page 10, algorithm step 7). 
In particular, in \fdrtool the function $\fdr(y)$ is estimated nonparametrically
and modeled by a step-function.

In the present work we study the estimation of FDR using two 
continuous variants of threshold curves for $\fdr(y)$.
Specifically, we consider the half-normal decay (HND) model by
\cite{RS2008} and the beta-uniform mixture (BUM) model of \cite{PM03}.
A further motivation for our study is the recent comparison by 
\citep{Mur2010} who found that the discrete fdr function obtained by
the \fdrtool algorithm may lead to a bias and thus is open to further
improvement.

\section{Models for fdr threshold curves}

We now discuss two simple local FDR threshold functions $\fdr(y)$.  There are
two natural properties
for such a curve. First, the function should be monotonically 
decreasing, so the FDR values lead to the same ranking as the 
raw statistics $y$.  Second, on a $z$-score scale the 
shape of the curve should be sigmoidal
 ranging
from $\fdr(0)=1$ onwards to $\fdr(y \rightarrow \infty) = 0$.
The beta-uniform mixture (BUM) and the half-normal decay (HND)
model, as well as their generalizations, satisfy these criteria.

\begin{figure}[!h]
\begin{center}
\centerline{\includegraphics[scale=.58]{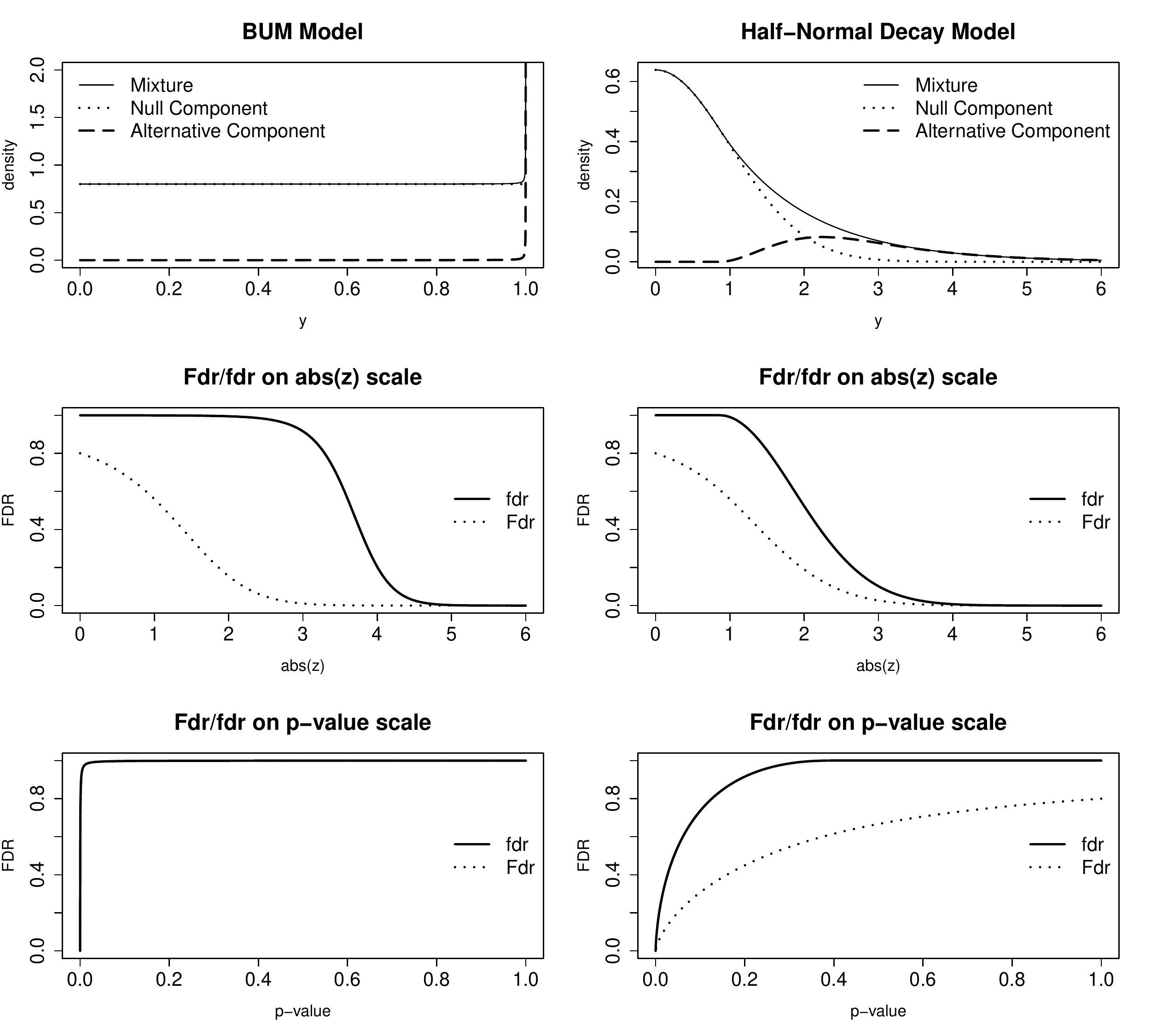}}
\caption{Examples of the BUM and HND model for $\eta_0=0.8$. The first
row shows the corresponding joint, null and alternative densities.
The second row displays the fdr and Fdr values on the standard normal 
$z$-score scale. The third row shows fdr and Fdr values on the $p$-value scale. }
\label{fig:bumhndmodel}
\end{center}
\end{figure}
\subsection{Beta-uniform mixture (BUM) model}

The BUM model was proposed in the context of FDR estimation
from $p$-values \citep{PM03}.  We define the model based on a random
variable $Y \in [0,1]$ with uniform distribution as null model.
The density is therefore
$$
f_0(y) = 1
$$
and the corresponding distribution
$$
F_0(y) = y \,. 
$$
The  BUM fdr function is given as a one parameter family
$$
\fdr^\text{BUM}(y | s)= \frac{s}{s + a (1-s) (1-y)^{a-1}} \,.
$$
Note that $a$ is not a parameter but a small constant so that
approximately $\fdr^\text{BUM}(0 | s) = 1$ (we use $a=0.001$
throughout).  From \eqcite{eq:eta0integral} we find
the identity
$$
\eta_0 = s
$$
which greatly facilitates the interpretation of the parameter $s$.
The marginal density in the BUM model is therefore (\eqcite{eq:marginaldist})
$$
f(y) = \eta_0 + a (1-\eta_0) (1-y)^{a-1} \,. 
$$
Similarly, the alternative density is
$$
f_A(y) = a (1-y)^{a-1}
$$
and the distribution
$$
F_A(y) = 1 - (1-y)^a \,.
$$
The resulting marginal distribution is
$$
F(y) = \eta_0 y + (1-\eta_0) ( 1 - (1-y)^a) 
$$
which leads with \eqcite{eq:tailareafdr}
to the following expression for the $q$-value
$$
\Fdr(y) = \frac{\eta_0}{\eta_0 + (1-\eta_0) (1-y)^{a-1} } 
$$
which has $\Fdr(0) = \eta_0$ as required.

The BUM can also be trivially reformulated using
p-values ($y(p)=1-p)$. Alternatively, as null statistic one may also 
use standard normal $z$-scores with $y(z) = 2 \Phi(|z|)-1 $
where $\Phi$ is the standard normal distribution function.
The Fdr and fdr curves are invariant against reparameterization,
i.e. $\Fdr(z) = \Fdr(y(z))$ and $\fdr(z) = \fdr(y(z))$.
The marginal density is computed as $f(z) = \eta_0 f_0(z)/\fdr(y(z))$ 
and thus requires as additional factor
the volume element (which is hidden here in the transformation from $f_0(y)$ to $f_0(z)$).

In \figcite{fig:bumhndmodel} the BUM model and the associated
Fdr and fdr values are shown for $\eta_0 = 0.8$ both on a $p$-value scale
and on a standard normal $z$-score scale.

\subsection{Half-normal decay (HND) model}

The half-normal decay model is first described by Rice and Spiegelhalter \cite{RS2008}.
Its starting point is the random variable $Y$ drawn from standard half-normal distribution
Thus, the observations $y \in [0, \infty]$ with null
density
$$
f_0(y) = \sqrt{\frac{2}{\pi}} e^{-y^2/2}
$$ 
and corresponding distribution function
$$
F_0(y) = 2 \Phi(y)-1 \, .
$$
The local FDR curve is given by a one parameter family
$$
\fdr^\text{HND}(y | s)=
\begin{cases}
1 & \text{for $y \leq s$} \\
e^{-(y-s)^2/2} & \text{for $y > s$} \, .
\end{cases}
$$
The parameter $s$ has a natural interpretation as cut-off threshold
below which there are no ``interesting'' cases.
This specification of null model and fdr curve results in
$$
\eta_0 = \biggl( 2 \Phi(s)-1 + \sqrt{\frac{2}{\pi}} e^{-s^2/2-\log s} \biggr)^{-1}
$$
This equation is invertible, hence the parameter $s$  has a one-to-one
correspondence to the proportion of the null features $\eta_0$.
In the HND model the marginal density is
$$
f(y)=
\begin{cases}
\eta_0 \sqrt{\frac{2}{\pi}} e^{-y^2/2} & \text{for $y \leq s$} \\
\eta_0 \sqrt{\frac{2}{\pi}} e^{s^2/2-y s} & \text{for $y > s$} \\
\end{cases}
$$
and the alternative density 
$$
f_A(y)=
\begin{cases}
0 & \text{for $y \leq s$} \\
\frac{\eta_0}{1-\eta_0} \sqrt{\frac{2}{\pi}} ( e^{s^2/2-y s} -e^{-y^2/2} ) & \text{for $y > s$} \,.\\
\end{cases}
$$ 
Finally, the marginal distribution function is
$$
F(y)=
\begin{cases}
\eta_0 (2 \Phi(y) -1 ) & \text{for $y \leq s$} \\
\eta_0 \biggl( 2 \Phi(s) -1 + \sqrt{\frac{2}{\pi}} e^{s^2/2-\log s} (e^{s^2}-e^{-s y}) \biggr) & \text{for $y > s$} \\
\end{cases}
$$
which together with $F_0(y)$ allows to compute the tail-area-based Fdr by applying \eqcite{eq:tailareafdr}.

The HND may also be expressed in terms of $p$-values, using the transformation $y = \Phi^{-1}(1-p/2)$.
In \figcite{fig:bumhndmodel} the HND model 
for $\eta_0=0.8$ (or equivalently $s=0.862$) is shown and contrasted with the notably
different BUM model.

\subsection{Generalizations and problem of confounding} 

The BUM and HND fdr threshold functions are one parameter families indexed
by the parameter $s$ which in both models has a one-to-one mapping onto 
the true  proportion of null hypotheses $\eta_0$.
To allow for more flexibility it is useful to introduce additional parameters,
either in the null density $f_0(y)$ or in the fdr function $\fdr(y)$. 
For example, if the null model is misspecified then an additional variance
 parameter is often all that is needed to extend the model \citep{Efr08a}.
On the other hand, if the alternative density is not flexible enough 
this may be fixed by introducing additional parameters into the fdr curve.

Here, we will employ both the BUM and HND model with an additional scale parameter $\sigma$
in the null model. Specifically, we assume that the null density is a normal 
$N(0,\sigma^2)$ with mean zero and variance $\sigma^2$ so that for the HND model $y=|z/\sigma|$ and for BUM $y=2\Phi(|z/\sigma|)-1$, where $z$ is
the observed test statistic.

In generalizing null models and fdr functions particular care is necessary because
of potential confounding of parameters, especially if the null model and the fdr threshold 
function are extended simultaneously.
For example, the local fdr curve of the standard 
HND model  has an inflection point at $z_0= s+1$ with fdr value  $e^{-1/2} \approx 0.6$ and slope $-e^{-1/2}\approx -0.6 $.   The extended HND model with additional scale
parameter $\sigma$ in the null model leads to an fdr curve with inflection point
$z_0 = s + \sigma$ with fdr value $-e^{-1/2} \approx 0.6$ and slope $-\sigma e^{-1/2}\approx -0.6 \sigma$.  Thus, the scale parameter of the null model directly determines the slope of the fdr curve  at its inflection point, which implies that scale  and slope parameters
are confounded.

\subsection{Empirical null} 

In a setting with a large number of multiple tests is is possible to 
employ an empirical null model \citep{Efr08a}.  Specifically, instead of 
assuming a theoretical null density with some fixed parameter $\sigma$ it is
  possible (and often beneficial) to estimate it from data .  As in the present framework the
marginal density is a  completely specified low-dimensional family given by the null
density and the fdr function (\eqcite{eq:marginaldist}), the empirical null can
 be obtained in a straightforward fashion by maximum likelihood estimation.

\section{Results}

We present results from the analysis of synthetic data, followed
by a reanalysis of four data sets from \citep{RS2008}.

\subsection{Setup of simulation study}

In order to evaluate the accuracy of the FDR threshold approach
we conducted computer simulations.
For the data generation we followed the simulation setup 
for $z$ scores described in \citep{Str08c}:
\begin{itemize}
\item
Data $x_1, \ldots, x_{200}$ were
drawn from a from a mixture of the normal distribution $N(\mu = 0, \sigma^2=4)$
with the symmetric uniform alternatives $U(-10, -5)$ and $U(5, 10)$ and a
null proportion of $\eta_0=0.8$. 
\item The sampling was repeated $B=1000$ times.
\end{itemize}
The alternative density of this model does not match the
implied alternative density $f_A$ of neither the BUM nor the HND parameterizations. 
Thus, with this simulation setup we investigate how well the fdr threshold model performs
under misspecification. Note that this mixture model corresponds to a scenario
where null and non-null features are  well separated.

In addition, we also simulated data for a scenario where the alternative
and the null model are overlapping:
\begin{itemize}
\item
Setup as above but with $U(-10, -2)$ and $U(2, 10)$ as alternative distribution.
\end{itemize}
This scenario leads to a marginal density that is similar in shape as the native HND model.

In the subsequent step of comparison of resulting FDR values
and model parameters we employed two different strategies for fitting the
parameters for the HND and BUM models
\begin{enumerate}
\item External estimation: the parameters of the fdr threshold model  $\sigma$ and $s$ (or equivalently $\eta_0$) are estimated using \fdrtool \citep{Str08b} and plugged
into the corresponding equations of the BUM and HND models. This allows to directly compare
the FDR values computed by \fdrtool with that of BUM and HND.
\item Empirical null model: the parameters of fdr threshold model are estimated by
maximizing the marginal likelihood of the BUM and HND models. We refer to these estimated
models as BUM-native and HND-native. This allows to evaluate the effect of misspecification
on parameter estimation.
\end{enumerate}
In each case we computed for all $B=1000$ repetitions the Fdr and fdr values
of all $m=200$ hypotheses and compared these
estimates with the true Fdr and fdr values as given by the true known mixture model.  

\subsection{Results from simulation study}

\begin{figure}[hp!]\centering

\subfloat[]
  {
     \includegraphics[width=\textwidth]{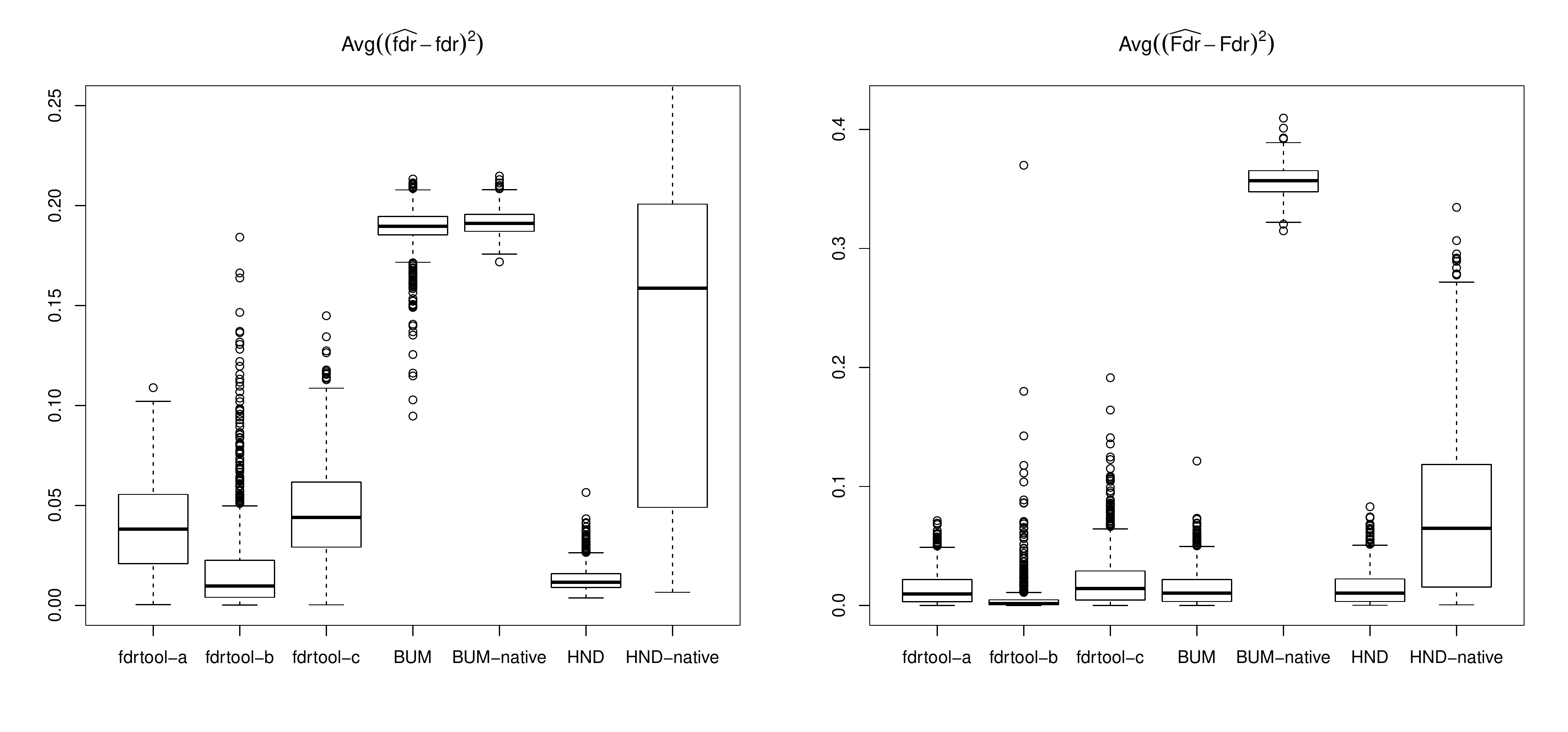}
     \label{fig:fdrFdrsim1}
  }

\subfloat[]
  {
     \includegraphics[width=\textwidth]{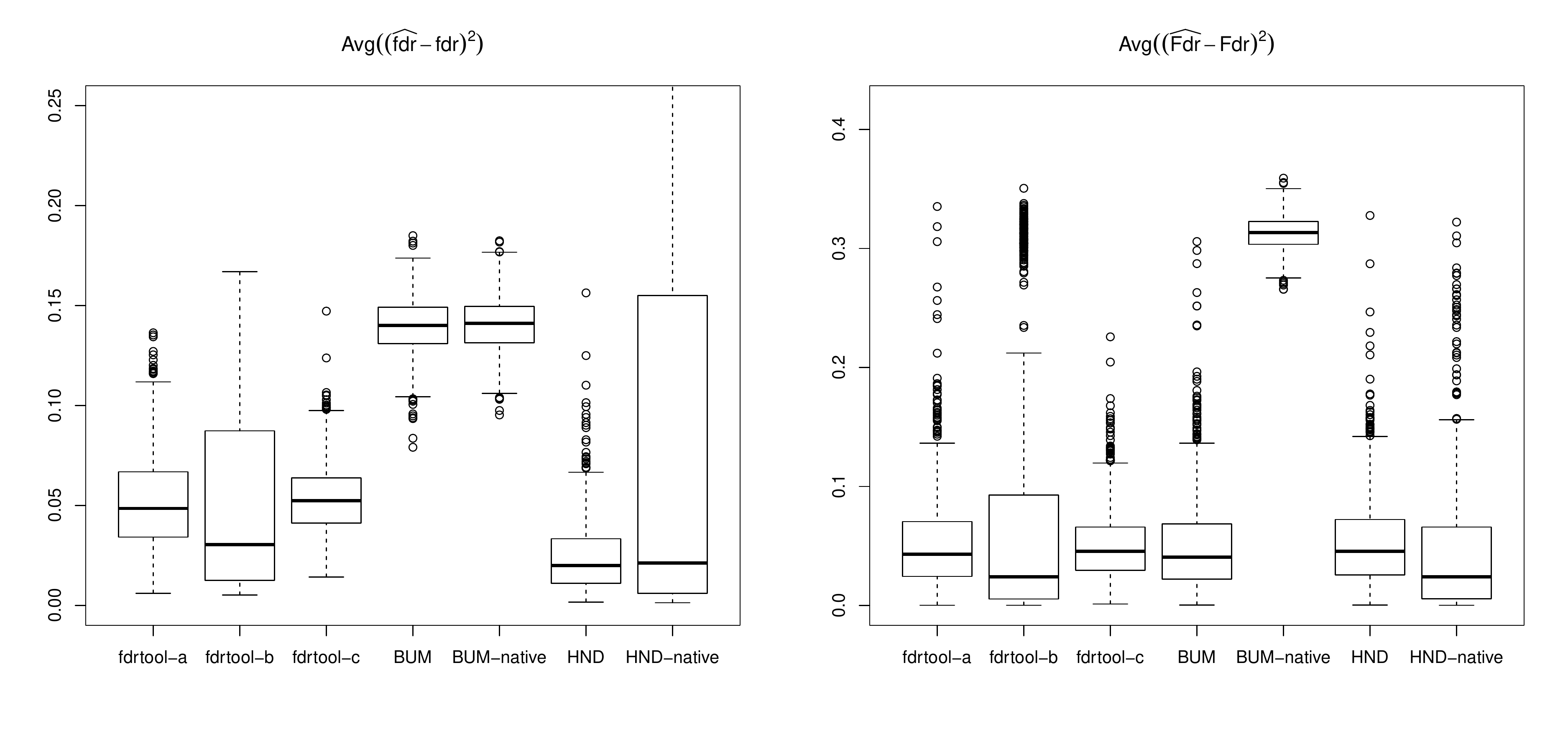}
     \label{fig:fdrFdrsim2}
  }

  \caption{\small Comparison of the accuracy of fdr and Fdr estimates for the simulated data:
        (a) well separated case, and
	(b) overlapping scenario.
	}

\label{fig:fdrFdrsim}
\end{figure}

In \figcite{fig:fdrFdrsim} the results from the comparison of true 
and estimated FDR values are shown using the following abbreviations
for the investigated algorithms: fdrtool-a, fdrtool-b, and fdrtool-c
correspond to using the \fdrtool software \citep{Str08b,Str08c}, with option
{\tt cutoff.method} set to {\tt fndr}, {\tt pct0}, and {\tt locfdr}, respectively
(note that fdrtool-a is the default method); BUM and HND denote the two fdr threshold
methods with null model given by fdrtool-a; and BUM-native and HND-native correspond
to the two fdr threshold
methods with empirical null model.

The results can be summarized as follows.  For local FDR (first column in \figcite{fig:fdrFdrsim}) the HND model improves over \fdrtool. As fdrtool-a uses exactly
the same null model as HND this shows that the step function used to model the
local FDR in \fdrtool may be improved by suitable smoothing.  Intriguingly, however,
HND-native exhibits a dramatic reduction of accuracy in fdr estimation if the null and 
alternative are well separated (upper left image). On the other hand, if the null and
the alternative are overlapping the HND-native approach performs well (albeit 
with a large variance). 
The BUM models performs worst, both with and without empirical null model.
For tail-area based FDR (second column in \figcite{fig:fdrFdrsim}) both BUM
and HND perform similar as \fdrtool.  However, there is again a drastic
reduction in accuracy for HND-native and BUM-native in the case of clear
separation of null and alternative density (upper right image). HND-native
performs very well in the overlapping scenario.

\begin{figure}[thp!]\centering

\subfloat[]
  {
     \includegraphics[width=\textwidth]{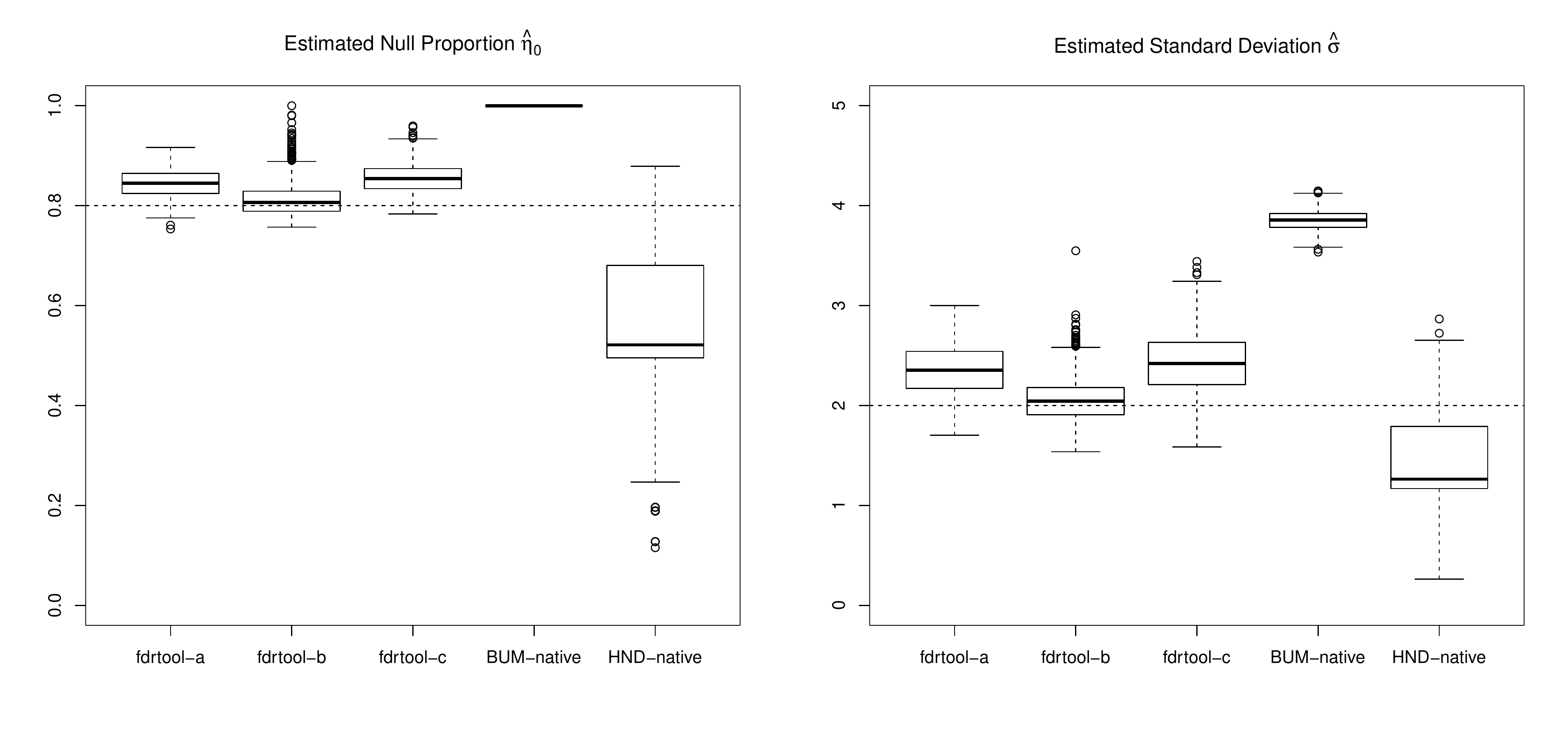}
     \label{fig:eta0sdsim1}
  }

\subfloat[]
  {
     \includegraphics[width=\textwidth]{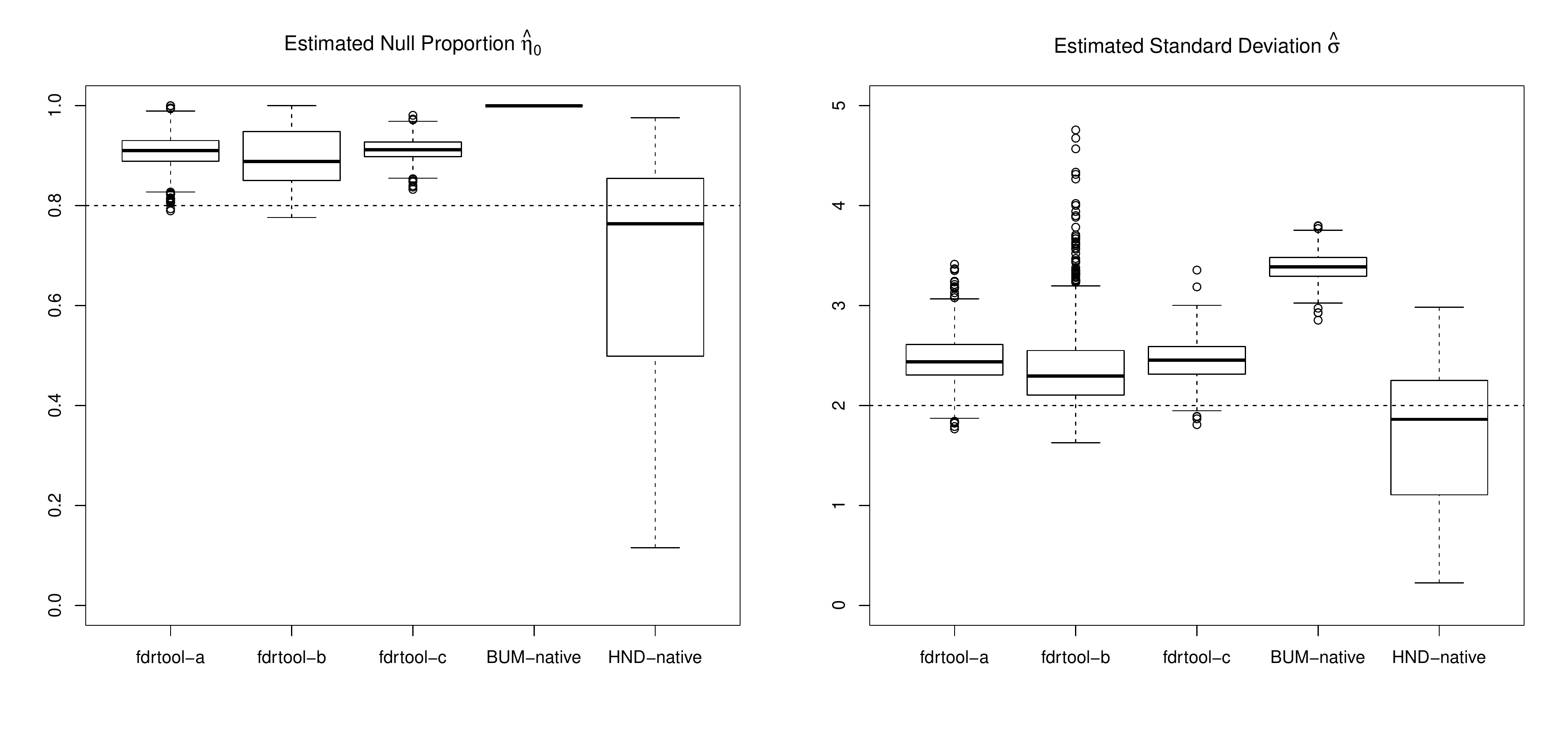}
     \label{fig:eta0sdsim2}
  }

  \caption{\small Comparison of the accuracy of parameter estimates for the simulated data:
        a) well separated case, and
	(b) overlapping scenario.
	}

\label{fig:eta0sdsim}
\end{figure}

\figcite{fig:eta0sdsim} shows the accuracy of the estimated
null models for fdrtool-a, fdrtool-b, fdrtool-c, BUM-native,
and HND-native.  In the first column box-plots for the estimated
null proportion $\hat{\eta_0}$ are shown. With the true value
of $\eta_0 = 0.8$ it is evident that BUM-native always 
overestimates $\eta_0$ whereas HND-native mostly underestimates
$\eta_0$.  Likewise, the second column shows that the scale parameter $\sigma$ is also always
overestimated by BUM-native and mostly underestimated by HND-native.
In comparison, \fdrtool overestimates both $\eta_0$ and $\sigma$ only slightly.
As in \figcite{fig:fdrFdrsim} we also clearly see the impact of the
misspecification on HND-native.  If the null and alternative densities
are clearly separated the HND-native model is not appropriate
but in the more difficult case of overlapping mixture components
HND-native performs rather well.

In summary, we find the HND model works well for both Fdr and fdr estimation
 if the correct parameters
for the null model are being supplied. HND-native estimation of the 
 empirical null requires that model and data are not misspecified.
In contrast, the BUM model is only suited for Fdr estimation and
empirical null estimation failed for both investigated scenarios.

\subsection{Analysis of real data}

\begin{table}[!t]
\caption{Parameter estimates obtained for four real data sets.}
\centering
\begin{tabular}{lcccc}
\toprule
             & Prostate  &  Education & BRCA & HIV \\
\midrule
$\hat\eta_0$:  \\
fdrtool-a    & 0.9855  & 0.9671  & 1      & 0.9587 \\
BUM-native   & 1       & 1       & 1      & 0.9984 \\
HND-native   & 0.9829  & 0.9536  & 1      &  0.9370 \\
\midrule
$\hat\sigma$: \\
fdrtool-a    & 1.0649  & 1.7204  & 1.5730 &  0.7999 \\
BUM-native   & 1.1350  & 1.9911  & 1.4313 &  0.9220 \\
HND-native   & 1.0588  & 1.6810  & 1.4311 &  0.7652 \\
\bottomrule
\end{tabular}
\label{tab:fourdatasets}\\

\end{table}

In their original paper Rice and Spiegelhalter  
analyzed for four experimental data sets 
concerning prostate cancer, education (mathematics competency),  
breast cancer and HIV \citep{RS2008}.  We refer to this paper
for details and biological background of the data.

\tabcite{tab:fourdatasets} shows the estimates of the model parameters
$\sigma$ and $\eta_0$ obtained by BUM-native and HND-native in comparison
with the  fdrtool-a algorithm.  In agreement with the simulations BUM-native
performs rather poorly, and HND-native underestimates relative to fdrtool-a.
However, in these data examples the fdrtool-a and HND-native are in broad 
agreement, which implies that here the implicit alternative density of the HND model
is appropriate.

\section{Conclusion}

FDR estimation by direct modeling the fdr threshold curve is a very elegant 
procedure.  We have investigated this procedure using two parametric models for
the fdr function and explored its robustness with respect to misspecification
of data and model with regard to estimation
of local and tail-area based FDR.

The original motivation for proposing this approach in \citep{RS2008}
was a preference of fully explicit modeling over using (supposedly) adhoc approaches.
However, as our study shows, a full specified model such as HND, and even more so BUM, runs
a severe risk of misspecification.  In such a case, semi- or nonparametric approaches such as
\citep{Str08c} or  \citep{Mur2010} are in our view preferable, especially if the number
of hypotheses is large.

\section*{Acknowledgments}
We thank Prof. David Spiegelhalter for kindly making available to us the four data sets
and Katja R\"osch for critical reading of the manuscript. We also thank the anonymous referee
for very helpful comments.
Part of this work was supported by BMBF grant  no. 0315452A (HaematoSys project).

%\newpage

\bibliographystyle{apalike}
\bibliography{preamble,econ,genome,stats,array,sysbio,misc,molevol,med,entropy}

\end{document}